\documentclass[10pt,twocolumn,letterpaper]{article}

\usepackage{cvpr}             
\usepackage{graphicx}
\usepackage{amsmath}
\usepackage{amssymb}
\usepackage{booktabs}

\usepackage{pifont}
\newcommand{\cmark}{\ding{51}}%
\newcommand{\xmark}{\ding{55}}%

\usepackage[pagebackref,breaklinks,colorlinks]{hyperref}

\usepackage{booktabs}
\usepackage[table,xcdraw]{xcolor}

\usepackage[capitalize]{cleveref}
\crefname{section}{Sec.}{Secs.}
\Crefname{section}{Section}{Sections}
\Crefname{table}{Table}{Tables}
\crefname{table}{Tab.}{Tabs.}
\begin{document}
\title{Quality Assessment of Low Light Restored Images: A Subjective Study and an Unsupervised Model}
\author{Vignesh Kannan, Sameer Malik, Rajiv Soundararajan\\
Indian Institute of Science\\
{\tt\small vigneshkanna@iisc.ac.in,  sameer@iisc.ac.in, rajivs@iisc.ac.in}
}
\maketitle
\begin{abstract}
  The quality assessment (QA) of restored low light images is an important tool for benchmarking and improving low light restoration (LLR) algorithms. While several LLR algorithms exist, the subjective perception of the restored images has been much less studied. Challenges in capturing aligned low light and well-lit image pairs and collecting a large number of human opinion scores of quality for training, warrant the design of unsupervised (or opinion unaware) no-reference (NR) QA methods. This work studies the subjective perception of low light restored images and their unsupervised NR QA. Our contributions are two-fold. We first create a dataset of restored low light images using various LLR methods, conduct a subjective QA study and benchmark the performance of existing QA methods. We then present a self-supervised contrastive learning technique to extract distortion aware features from the restored low light images. We show that these features can be effectively used to build an opinion unaware image quality analyzer. Detailed experiments reveal that our unsupervised NR QA model achieves state-of-the-art performance among all such quality measures for low light restored images. 

\end{abstract}

\section{Introduction}
\label{sec:intro}
The ability to produce good quality images under low ambient lighting conditions is an important feature of modern mobile photography. Low light image restoration involves solving the twin challenges of contrast enhancement and denoising. While several classical approaches based on histogram equalization \cite{WAHE,CVC,CLAHE,LDR} and retinex theory \cite{LIME,Robustretinex} have been studied in literature, the emergence of deep learning techniques \cite{Lv2018MBLLENLI,SID,Vu2018FastAE,Yu2019DeepID,DBLP:journals/corr/abs-1905-04161} has evoked a lot of research interest in this area. Despite the existence of a plethora of low light restoration algorithms, the quality assessment (QA) of the restored images has received much less attention. The focus of our work is in the study of perceptual quality assessment of low light restored images. 

Although several large scale generic image QA (IQA) databases exist, the subjective quality assessment of low light images is still a nascent research area. The camera image database (CID) \cite{DBLP:journals/tip/VirtanenNVOH15} consists of a variety of images captured by cameras, including low light distorted images. The natural night-time image database (NNID) \cite{DBLP:journals/tmm/XiangYG20} consists of a diverse set of low light images captured from three cameras. While the subjective QA of low light image captures has been studied, there is no large-scale study of low light  restored images. The distortions that arise while applying contrast enhancement and denoising algorithms on low light images are more diverse and different from images captured under low light conditions. The problems associated with the restoration algorithms themselves lead to several artifacts. In particular, low light restored images often suffer from various combinations of blur, noise, over/under enhancement, poor color saturation, and color casts. Thus there is a need for a detailed subjective study of low light restored images. This would help better compare the performance of low light images restored using different algorithms and advance their design. 

On the other hand, the objective assessment of low light restored images is also less explored. While full reference algorithms such as peak signal to noise ratio (PSNR), structural similarity (SSIM) \cite{DBLP:journals/tip/WangBSS04} and LPIPS \cite{DBLP:journals/corr/abs-1801-03924} are typically used for assessing the quality of the low light restored images, the requirement for a reference high quality image can be a limitation. Thus, there is a need to design no reference (NR) IQA algorithms. Further, NR IQA algorithms are often designed by training on a dataset of human opinion scores and evaluated on a test set. Since it is cumbersome to collect such human scores, such supervised algorithms are difficult to train on each dataset. Their generalization performance across datasets has also been limited \cite{6272356}. Thus, there is a need to study unsupervised NR IQA algorithms without any training on human opinion scores. 

The NIQE \cite{DBLP:journals/spl/MittalSB13} and IL-NIQE \cite{7094273} indices represent seminal examples of unsupervised NR IQA algorithms. They are designed by extracting natural scene statistics based features and comparing with a corpus of pristine images. Since, they do not particularly account for the distortions that arise during low light restoration, our experiments reveal that their performance for evaluating low light restored images is limited. This motivates the study and design of unsupervised NR IQA algorithms for low light restored images. Our goal is to design NR IQA algorithms that use a large corpus of low light restored images without training them using human labels. 

Our main contributions in addressing the above challenges are as follows:

\textbf{Dataset for Subjective Assessment of Low Light Restoration.} We create a large dataset of 1035 low light restored images obtained from a variety of contrast enhancement, denoising and joint contrast enhancement and denoising algorithms. Our dataset is unique and different from other IQA datasets in terms of the generation of distorted images through low light restoration algorithms. We then conduct an online subjective study involving 88 human subjects with 22,500 ratings. 

\textbf{Performance Benchmarking of IQA Measures on Low Light Restored Images.} We conduct a detailed benchmarking of the performance of several popularly used full reference and no reference image quality measures against the subjective scores obtained through our study. 

\textbf{Unsupervised Quality Assessment of Low Light Restored Images.} We present a scene wise multi-scale multi-view contrastive learning method on a large corpus of low light restored images to learn quality aware features. We then compare these features against features from a corpus of unpaired sharp and colorful high quality images captured under low light conditions to design an unsupervised NR IQA model. We refer to our method as M-SCQALE (Multiscale - Sub-band Contrastive learning for Quality Assessment of Low light Enhancement). We show that M-SCQALE achieves the state-of-the-art performance in terms of correlation with human scores when compared to all other NR unsupervised quality measures.

Our database and code will be made publicly available.

\section{Related work}
\textbf{Datasets for QA of low light restoration:}
While low-light restoration datasets such as SID \cite{SID}, LOL \cite{DBLP:conf/bmvc/WeiWY018}, ELD \cite{wei2020physics} contaning low-light / well-lit pairs exist in literature, there are no QA studies of low light restored images. Xiang \textit{\etal} \cite{DBLP:journals/tmm/XiangYG20} create a night time IQA dataset which assesses low light images having camera/device induced distortions. CID \cite{DBLP:journals/tip/VirtanenNVOH15} is an older database containing fewer images, a more diverse set of distortions but catering to both low light and well lit scenarios. Both the above datasets contain camera captured distortions but do not address images obtained through low light restoration algorithms. 

\textbf{No Reference image quality assessment:}
NR-IQA methods can broadly be divided into two categories viz. supervised (or opinion aware) and unsupervised (or opinion unaware) methods. 
Classical supervised method involve designs of handcrafted features which are regressed to learn the mapping to the quality labels. Some popular methods include BRISQUE, an NSS based method that extracts features in the spatial domain \cite{6272356} and CORNIA, a dictionary based feature learning framework \cite{CORNIA}. 
Several deep learning methods have also been successfully used for NR IQA. Zhang \etal \cite{2020} propose using a deep bilinear network that works well for both synthetic as well as authentic distortions. Zhu \etal \cite{Zhu2020MetaIQA} propose a meta-learning based method to learn a quality prior model which is then fine tuned on the target NR-IQA task. Other approaches include learning the distribution of scores instead of a single label \cite{Talebi2018NIMANI} and studying the relationships between patches and picture quality \cite{DBLP:journals/corr/abs-1912-10088}.
Among methods to assess low light camera captured distortions, Xiang \etal \cite{DBLP:journals/tmm/XiangYG20}  propose a night time IQA metric learnt using brightness and texture features. 

Unsupervised methods are much less explored in literature. Mittal \etal \cite{DBLP:journals/spl/MittalSB13} propose a completely blind quality analyzer that is based on measuring deviations from statistical regularities found in natural images. Zhang \etal \cite{7094273} extend this frame to enrich features with more statistical features. 
Gu \etal \cite{DBLP:journals/corr/abs-1904-08879} design a contrast distortion based metric which combines local as well as global aspects of an image. Nevertheless, unsupervised quality measures have not been specifically studied in the context of low light image restoration. 

\textbf{Self-Supervised feature learning:} 
Self-supervised learning has been studied extensively in literature for image classification, object detection and segmentation. Among these, contrastive learning based methods \cite{DBLP:journals/corr/abs-2006-10029,DBLP:journals/corr/abs-2003-04297,DBLP:journals/corr/abs-1906-05849,DBLP:journals/corr/abs-2006-09882,Wu2018UnsupervisedFL} have been very successful. Wu \etal \cite{Wu2018UnsupervisedFL} propose an instance discrimination based classification problem and use noise contrastive estimation to learn features. He \etal \cite{DBLP:journals/corr/abs-1911-05722} build large and consistent dynamic dictionaries for unsupervised feature learning. Chen \etal \cite{DBLP:journals/corr/abs-2002-05709} do away with the need for memory banks and specialized architectures by proposing a simple framework for contrastive learning of visually relevant features. Tian \etal \cite{DBLP:journals/corr/abs-1906-05849} present a multiview contrastive learning method such that the mutual information between different views of a scene is maximized. We find that the multiview contrastive learning approach lends itself naturally to isolate the quality aware features.

\section{Dataset and Subjective Study}
\label{sec:Creation of dataset and subjective study}
 We create \textbf{DSLR - Dataset for Subjective assessment of Low light Restoration}. DSLR is unique in terms of the application of diverse low light restoration algorithms on low light captured images leading to a wide spectrum of distortion types. It consists of 1035 low light restored images across 80 different scenes. Further we conduct an online subjective study thereby collecting a total of 22,500 quality ratings from 88 subjects. 

%-------------------------------------------------------------------------
\subsection{Dataset}

To create the low light restored images, we use low-light images from two publicly available datasets SID (test set) \cite{SID} and ELD \cite{wei2020physics}. These images are converted from RAW format to sRGB using the python rawpy library. Further we downsample the images to a resolution of 624 $\times$ 936 without changing its aspect ratio. We use a combination of 14 contrast enhancement techniques coupled with 3 denoising methods, and also 7 joint contrast enhancement and denoising techniques including deep learning methods to generate a huge corpus of around 27,000 images. A detailed list of techniques used can be found in the supplementary.

\begin{figure*}
\centering
\begin{subfigure}{0.23\linewidth}
    \centering
    \includegraphics[width=0.95\linewidth]{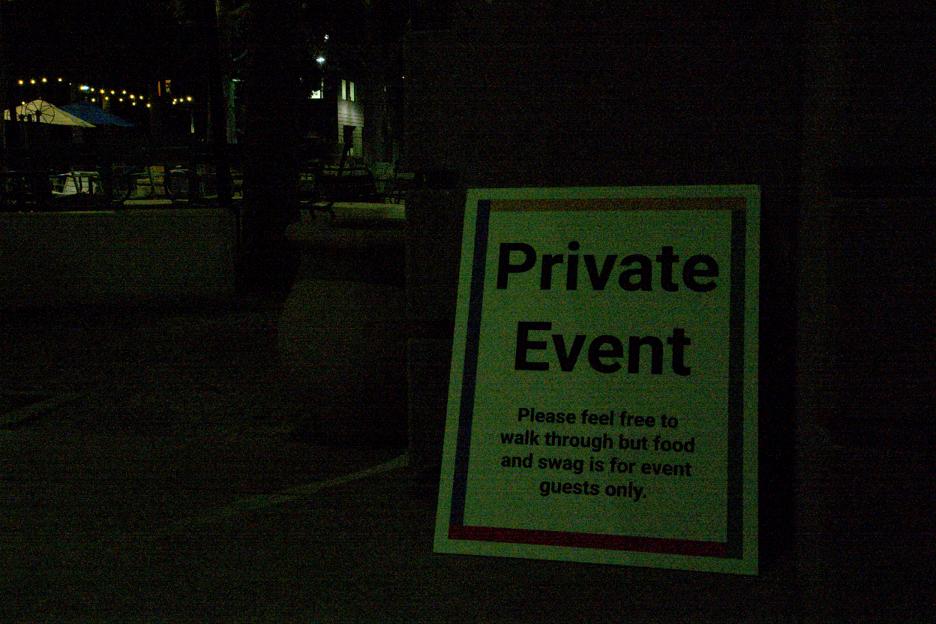}
    \caption{Low light}
\end{subfigure}
\begin{subfigure}{0.23\linewidth}
    \centering
    \includegraphics[width=0.95\linewidth]{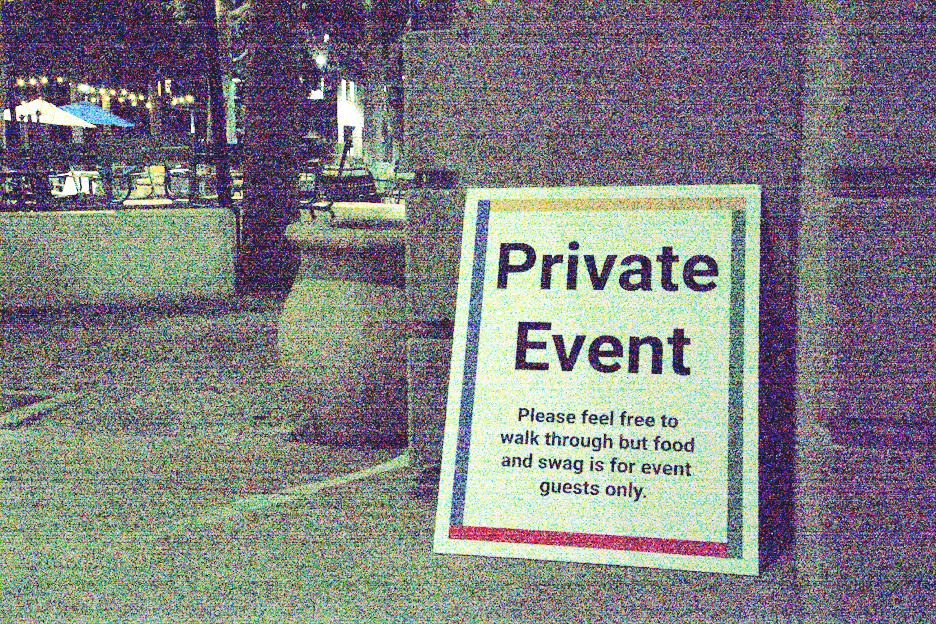}
    \caption{Noise}
\end{subfigure}
\begin{subfigure}{.23\textwidth}
    \centering
    \includegraphics[width=0.95\textwidth]{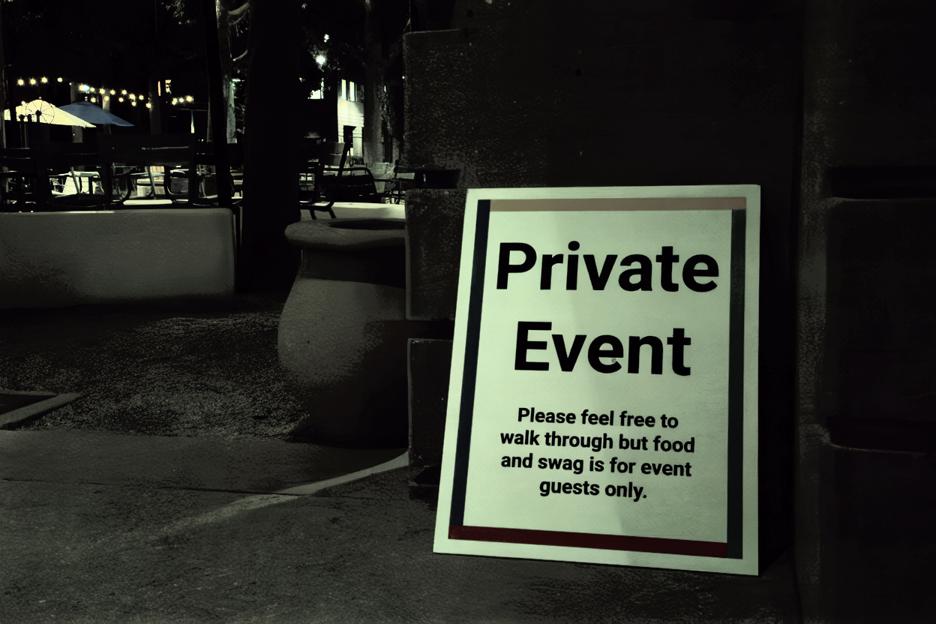}
    \caption{Poor color saturation}
\end{subfigure}
\begin{subfigure}{.23\textwidth}
    \centering
    \includegraphics[width=0.95\textwidth]{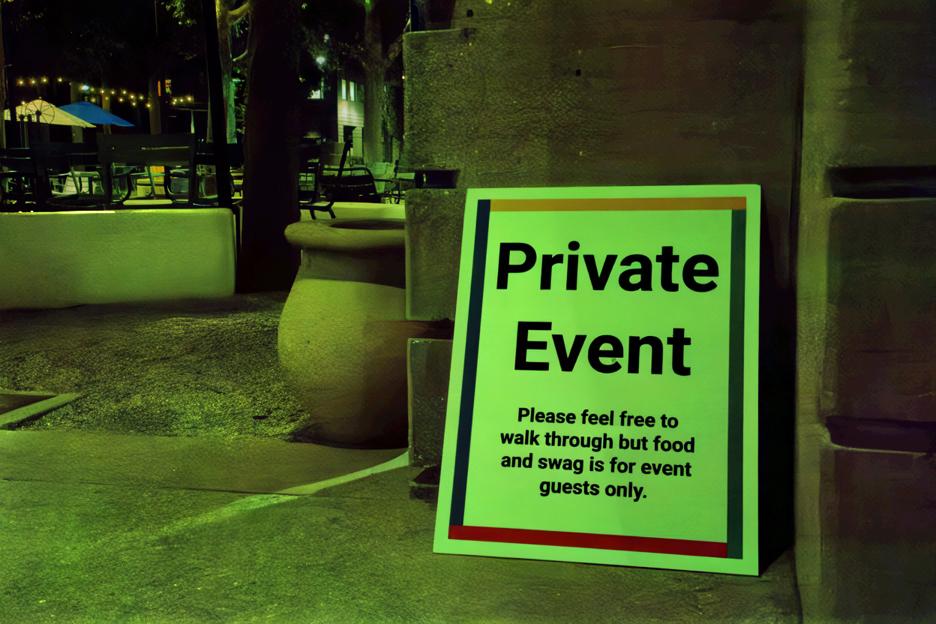}
    \caption{Colorcast}
\end{subfigure}
\begin{subfigure}{0.23\linewidth}
    \centering
    \includegraphics[width=0.95\linewidth]{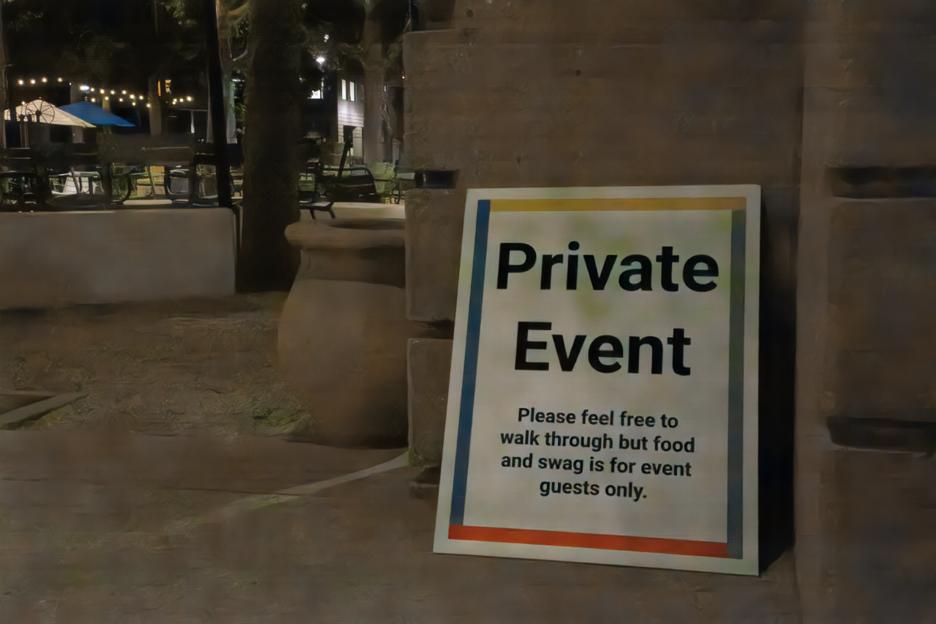}
    \caption{Denoise Blur}
\end{subfigure}
\begin{subfigure}{0.23\linewidth}
    \centering
    \includegraphics[width=0.95\linewidth]{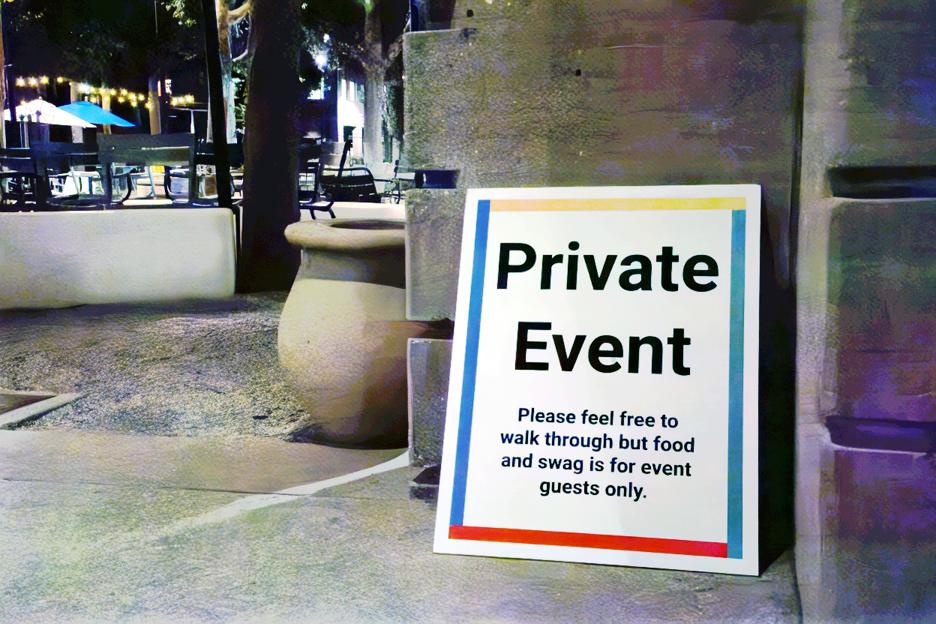}
    \caption{Over-enhancement}
\end{subfigure}
\begin{subfigure}{.23\textwidth}
    \centering
    \includegraphics[width=0.95\textwidth]{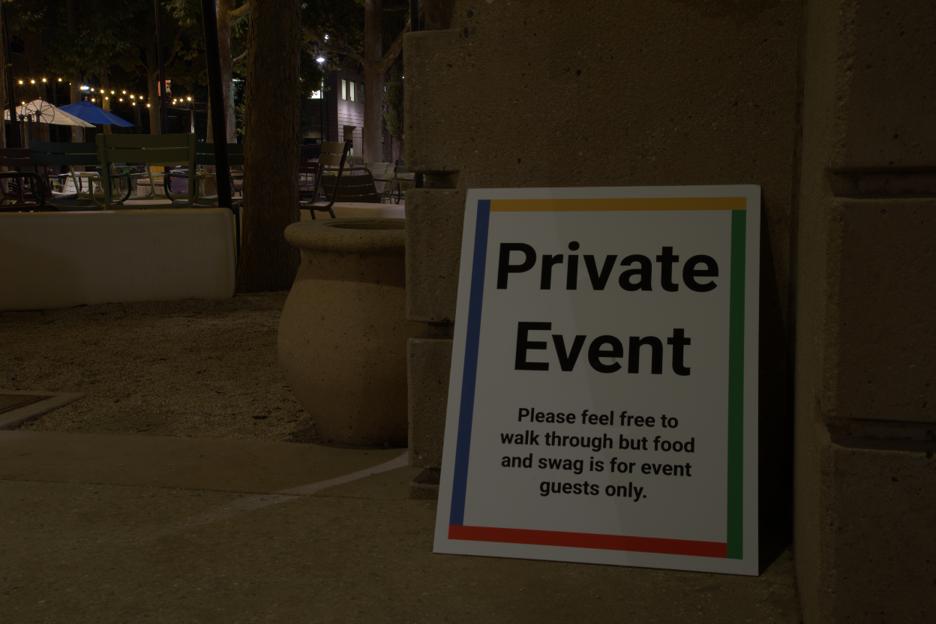}
    \caption{Under-enhancement}
\end{subfigure}
\begin{subfigure}{.23\textwidth}
    \centering
    \includegraphics[width=0.95\textwidth]{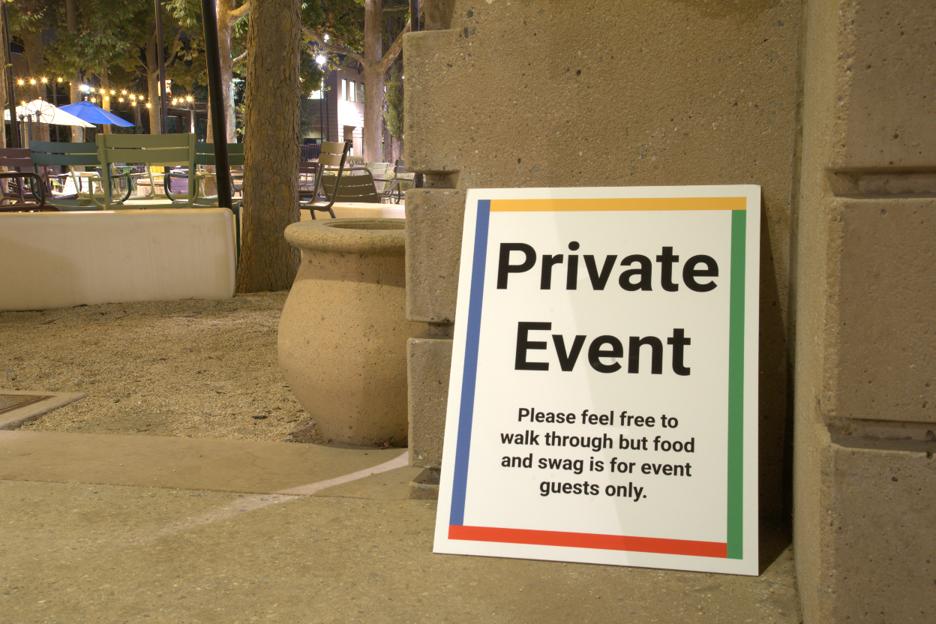}
    \caption{Well-lit}
\end{subfigure}
\caption{Low light, well-lit and different restored versions of a low light image showing various types of distortions in the dataset.}
\label{fig:distortion_types}
\end{figure*}

Since there is redundancy in the nature of distortions in this large corpus, we select a subset of 1035 images for the subjective study. We perform this selection by first identifying majorly occurring distortion types such as noise, blur, color distortions, over-enhancement, under-enhancement, and poor color saturation. For distorted versions of each scene, we classify the images according to the above distortion types, and select a subset of images for these distortion types spanning the low, medium and high quality ranges. When a restored image suffers from multiple distortion types, we associate it with the dominant distortion in the above selection steps. While finalizing on the set of distortions for a particular scene we make sure that the low light as well as the high exposure version of the scene available in the respective datasets \cite{SID,wei2020physics} is picked. We call the high exposure version of the scene as `well-lit'. We also include all the images from other distortion types such as halo artifacts, streak noise, fog which were few in number. In Figure \ref{fig:distortion_types}, we show a low light image and corresponding restored versions suffering from different distortion types.

%-------------------------------------------------------------------------

\subsection{Subjective study}
We conduct an online subjective study owing to the pandemic situation where each subject participates in two sessions of half an hour each spaced 24 hours apart. 
The study protocols were reviewed and approved by an appropriate committee overseeing human ethics. 
The human subjects were informed about the purpose of the study and consent obtained about how their data would be used.
Neither the study nor the database will reveal personally identifiable information. 

Although the study is done online where users rated images under different conditions similar to other crowdsourced studies \cite{Ghadiyaram2016MassiveOC,DBLP:journals/corr/abs-1912-10088,DBLP:journals/corr/abs-1910-06180,EvaluatingAM,Sinno2019LargeScaleSO}, we observe that there is a high level of agreement among the subjects as seen in Section \ref{sub_study_inf}. A total of 88 subjects took our study leading to around 20 or more ratings per image.

 We conduct a single stimulus continuous procedure study, where the subjects were asked to rate an image based on its perceptual quality on a continuous scale from 0-100. The entire study interface was designed and implemented in-house using a Node.js + MySQL backend. Each viewing session was preceded by a training phase so that the subjects get accustomed to the framework as well as get a sense of the range of qualities that the study might span. 
 Within a session, images were shown in a randomized order, taking care that no two images from the same scene occur in succession. The subject was allowed to go to the next image only after selecting a value on the slider. 

\textbf{Processing of subjective scores:}
We first convert raw human scores  to `Z-scores' \cite{LIVEIQA} by removing the mean and dividing by the standard deviation for each subject for each session. We then employ standard outlier rejection procedures described in \cite{itu2002methodology}. A total of 5 subjects were discarded as outliers. Finally the scores from the inlier subjects are linearly rescaled to the range 0-100 to obtain the final MOS scores. 

\subsection{Subjective Study Inferences}\label{sub_study_inf}

\begin{figure}
\centering
          \includegraphics[width=\linewidth]{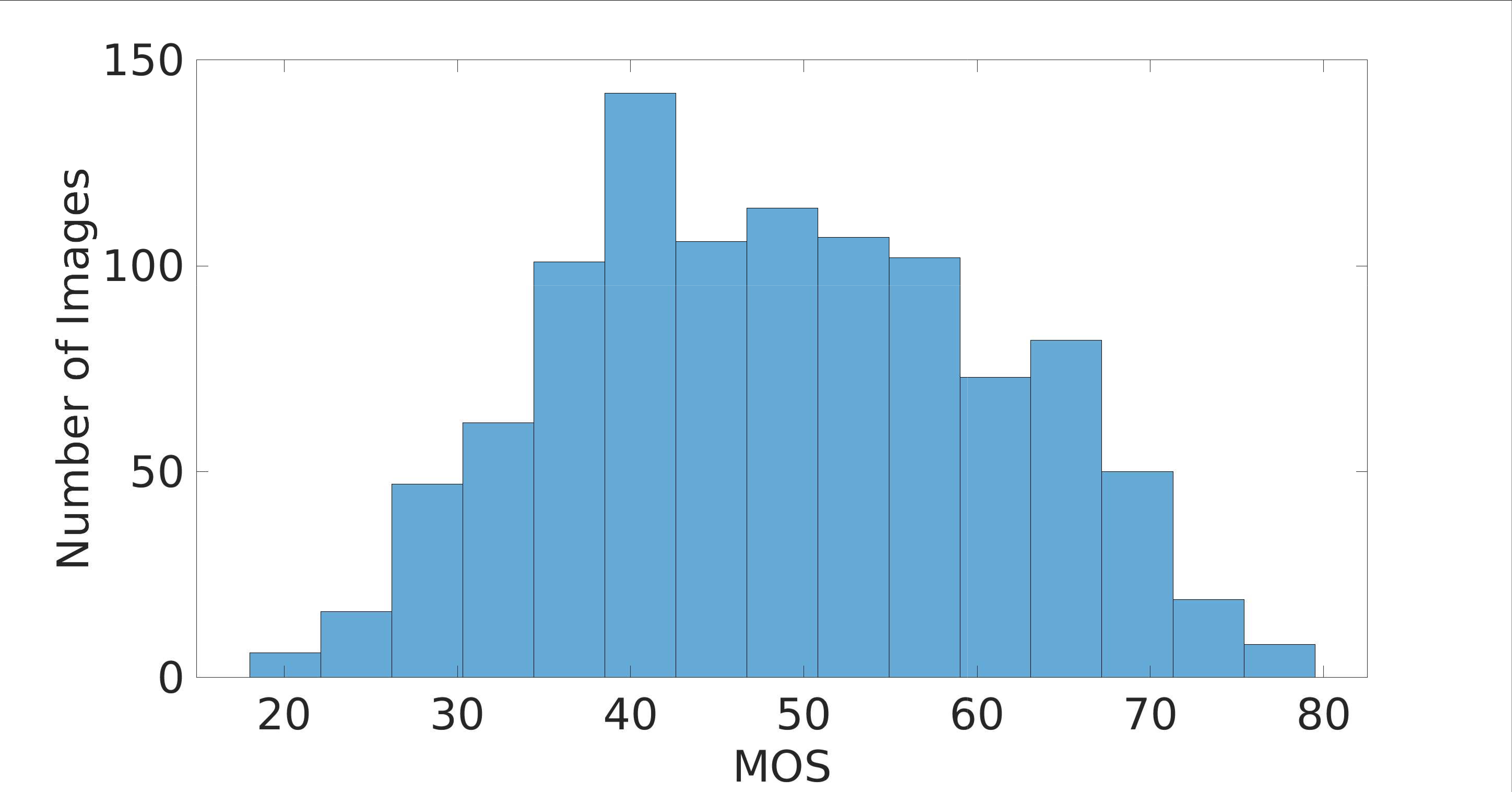}

          \caption{Distribution of MOS values across dataset.}
          \label{fig:MOS curve}
    \end{figure}

\textbf{MOS Distribution:} We show the MOS distribution across the dataset in Figure \ref{fig:MOS curve}. We see that the MOS values span the entire subjective quality range with more number of medium quality images as desirable in any challenging IQA database. 

%-------------------------------------------------------------------------

\begin{figure}
        \centering
          \includegraphics[width=\linewidth]{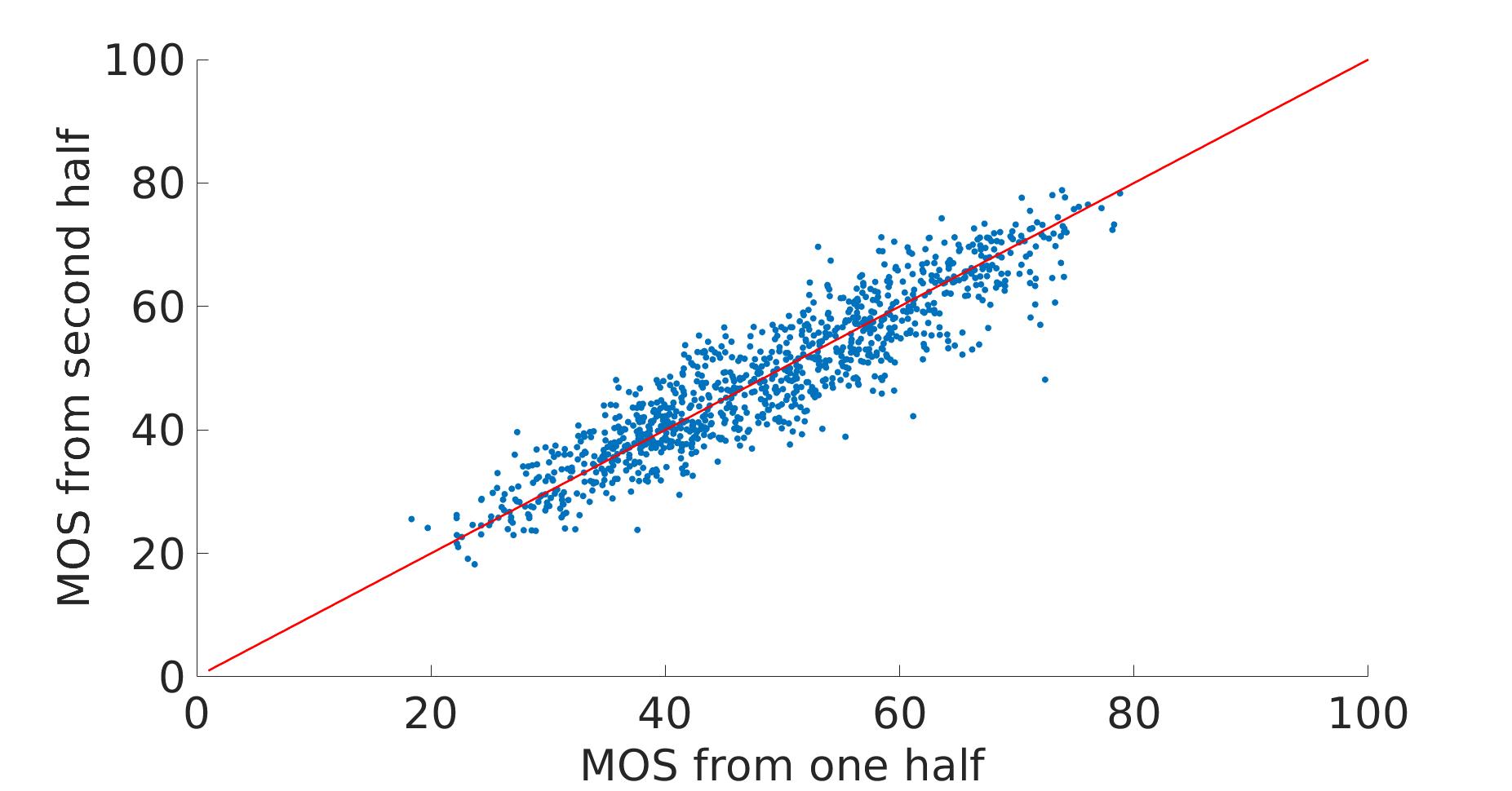}
              \caption{Inter-subject consistency between two random halves of the population.}
              \label{fig:MOS scatter}
        \end{figure}

    \textbf{Inter-subject consistency:}
    We validate the study by randomly splitting the population into two equal halves and compute the median Pearson linear correlation coefficient (PLCC) between the MOS obtained from the two halves across a number of random splits. We find that the median PLCC across 100 such splits is 0.93. We show the scatter plot for one such split in Figure \ref{fig:MOS scatter}. We see that there is a high agreement between two random halves of the population.
    
     \textbf{Validation using lab controlled study:}
    In order to validate our online study, we also conduct an in-lab study with controlled conditions. We use a 24 inch LED monitor with a screen resolution of 1920x1080. The subjects are positioned at a viewing distance of approximately four times the screen height. The room illumination is 6 lux. These settings are decided based on the recommendations given in \cite{itu2002methodology}. We employ a limited number of 10 subjects in our in-lab study. Each subject participated in a single session of half an hour each and was shown the same set of 166 images from DSLR. The PLCC between the MOS scores computed from the online study vs the in-lab study is 0.89. This validates the authenticity of the online study. 
  
\begin{figure}
    \centering
          \includegraphics[width=\linewidth]{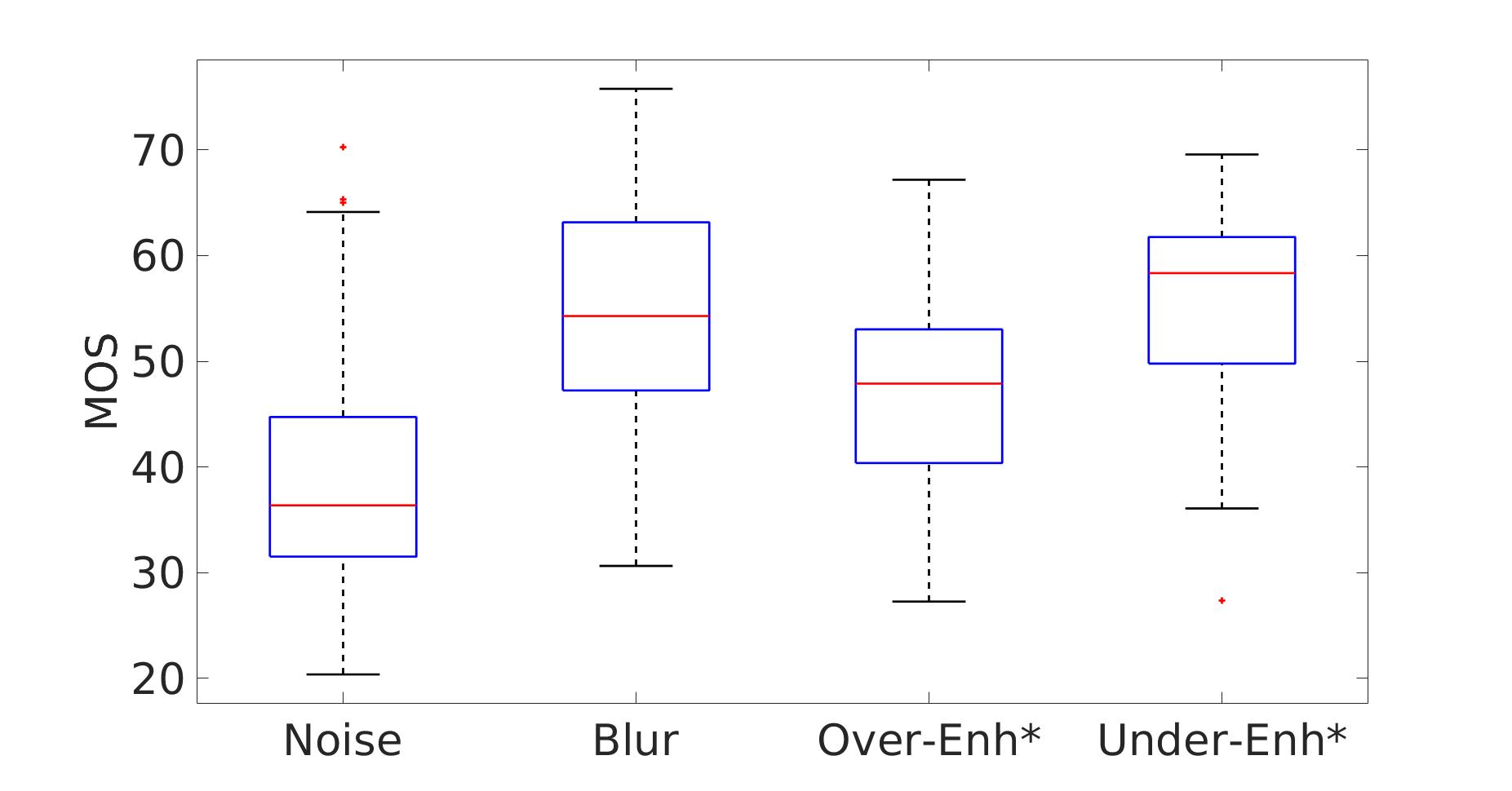}
          \caption{Boxplot of MOS values for different distortions. Here Enh* refers to Enhancement}
          \label{fig:distortion-wise boxplots}
    \end{figure}

    \textbf{MOS variability across distortions:}
     The MOS values of all images in our dataset lie in the range [20.36,78.26]. We note that the high exposure shots have MOS scores towards the higher end of the spectrum with a mean of 67.94 $\pm$ 5.65. The low light images have MOS scores towards the lower end of the spectrum with a mean of 33.45 $\pm$ 6.66. Figure \ref{fig:distortion-wise boxplots} shows box plots for different distortion types. As each comparative distortion type has roughly an equal span over the entire quality range, we draw certain conclusions from the boxplots in Figure \ref{fig:distortion-wise boxplots}. We observe that humans find noise more annoying when compared to blur, and over-enhancement more annoying when compared to under-enhancement. We also conduct two sample t-tests to validate the statistical significance of the above conclusions. 
   
\section{Unsupervised NR Low Light Restored IQA}
We now present the details of M-SCQALE, our unsupervised NR IQA method for low light restored images. We first present the self-supervised quality feature learning approach followed by the quality computation method.
\subsection{Contrastive Learning of Quality Features}
We present our approach for self-supervised learning of quality features for low light image restoration through multi-view contrastive learning \cite{DBLP:journals/corr/abs-1906-05849}. One of our key contributions is in the choice of the multiple views as relevant for learning quality features. We hypothesize that the joint distribution of quality features among different patches from the same distorted image can be contrasted with patches drawn from different distorted images to learn rich features. In particular, we divide a low light restored image into  non-overlapping patches and choose different patches as multiple views. 
Secondly, our choice of positive and negative pairs enables the learning of features that are sensitive to quality and less variant to content. 

In particular, we draw the positive and negative pairs of  views from distorted versions of the same scene in a manner that enables the learning of quality relevant features. Here, the only difference among the positive and negative pairs is in the visual quality while the content remains the same. We learn these features in each sub-band of a multi-scale Laplacian pyramid decomposition to mimic the multi-scale processing of human vision \cite{1292216}. 

Through our choice of patches in contrastive multi-view coding, the joint distribution of patches drawn from the same image tends to capture global quality features that are shared across the image. We believe that such feature extraction is analogous to how global features are extracted by modeling the natural scene statistics of image sub-bands for QA. Such feature extraction helps obtain a global sense of the image quality which is ultimately predicted as a single number. The global features are still a function of local features obtained through several layers of convolutions and non-linear activations in early stages. 

\begin{figure}
    \centering
          \includegraphics[width=\linewidth]{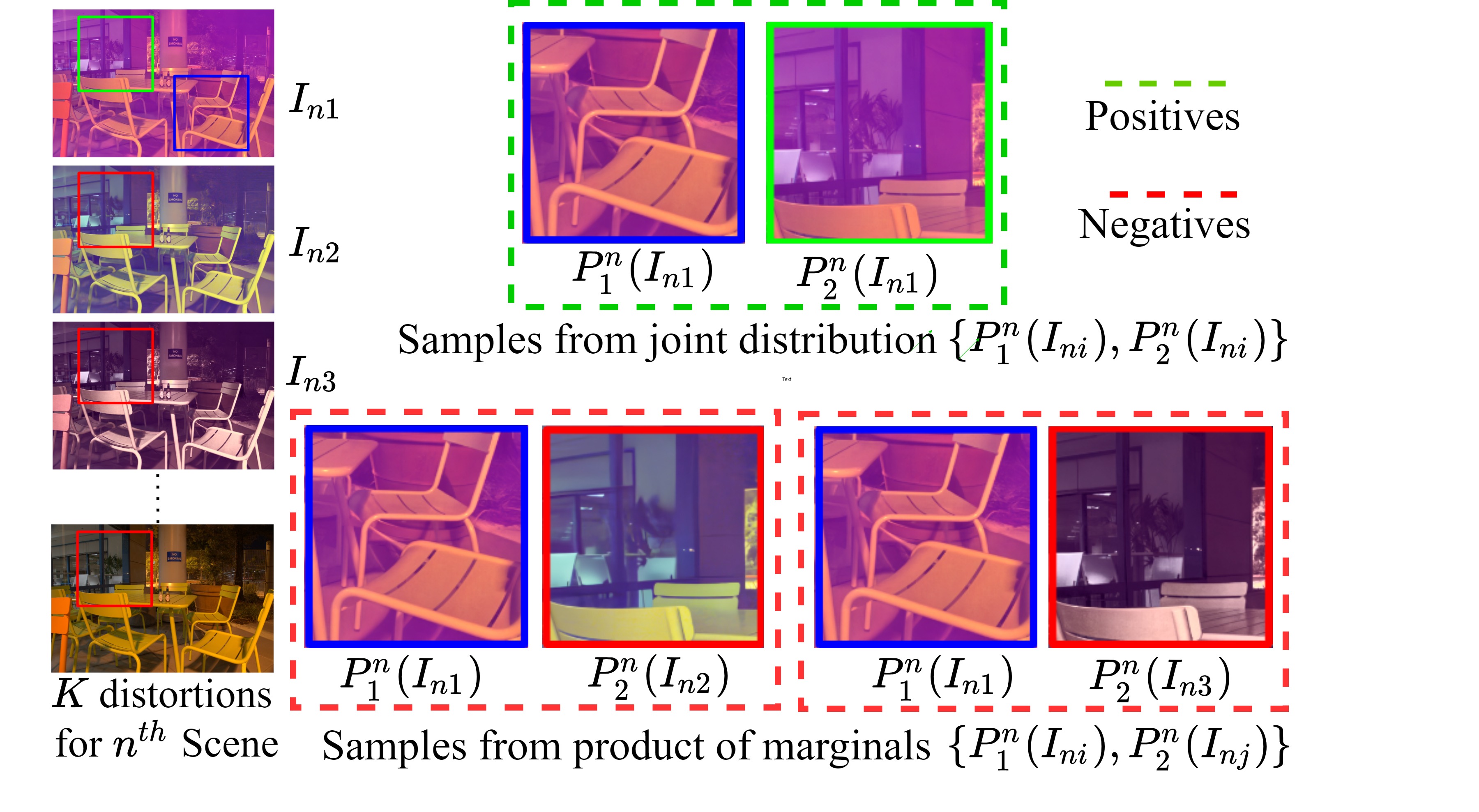}
          \caption{Illustration of views in our contrastive learning framework.}
          \label{fig:Our learning framework}
    \end{figure}

Now we formulate our multi-view contrastive learning as shown in Figure
\ref{fig:Our learning framework} and describe it below. For a given scene $n\in\{1,2,\cdots,N\}$, we choose $K$ distorted versions denoted as $\{I_{n1},I_{n2},I_{n3},\cdots,I_{nK}\}$. These distorted versions correspond to either the original low light image or its restored versions obtained by applying different contrast enhancement and denoising algorithms. 
We also include an image captured with high exposure as one of these distorted versions, although it may be of high quality. We now describe the feature learning from the images. The same procedure is repeated for each subband independently.
    
We define two functions $P^n_{1}(.)$ and
$P^n_{2}(.)$ where $P^n_{1}(.)$ crops a patch from the images in the batch  $\{I_{n1},I_{n21}...I_{nK} \}$ at the exactly same location while $P^n_{2}(.)$ corresponds to a different non-overlapping patch. For simplicity, we divide the image into vertical or horizontal halves and randomly pick the largest square patch in each half. 
We note that the patch location is the same across distorted versions of the same scene but can vary for different scenes. 
Thus, we obtain a total of $2K$ patches for each scene where View 1 contains $\{P^n_{1}(I_{n1}),\cdots,P^n_{1}(I_{nK})\}$ and View 2 contains $\{P^n_{2}(I_{n1}),\cdots,P^n_{2}(I_{nK})\}$.

We pick a random patch from View 1 and label it as an anchor. Let $P^n_{1}(I_{nk})$ be the anchor. Corresponding to this anchor, $(P^n_{1}(I_{nk}),P^n_2(I_{nk}))$ is a positive pair of samples since they are drawn from the same image while $(P^n_{1}(I_{nk}),P^n_2(I_{nj}))$ for $j\neq k$ is a negative pair of samples. Let the feature extraction network be denoted as $f(\cdot)$. Let $\mathbf{z}_{k1}^{(n)} = f(P_1^n(I_{nk}))$ and $\mathbf{z}_{k2}^{(n)} = f(P_2^n(I_{nk}))$. Let $S(\cdot)$ denote the normalized cosine similarity between two vectors $\mathbf{u}$ and $\mathbf{v}$ be defined as
\begin{equation*}
S(\mathbf{u},\mathbf{v}) = \frac{\mathbf{u}^T\mathbf{v}}{||\mathbf{u}|| ||\mathbf{v}||}.
\end{equation*}

The contrastive loss corresponding to anchor $P^n_{1}(I_{nk})$ is obtained as
\begin{equation}
    l(P^n_{1}(I_{nk})) =-\log \frac{\exp\left(S(\mathbf{z}_{k1}^{(n)},\mathbf{z}_{k2}^{(n)})/\tau \right)}{\sum_{j=1}^K\exp\left(S(\mathbf{z}_{k1}^{(n)},\mathbf{z}_{j2}^{(n)})/\tau\right)},
\end{equation}
where $\tau$ is the temperature parameter. We obtain the loss similarly when $P_2^n(I_{nk})$ is chosen as the anchor. Thus, the overall loss function for $N$ scenes in a batch is obtained as
\begin{equation}
    \mathcal{L} = \frac{1}{NK}\sum_{n=1}^N\sum_{k=1}^K \left[l(P_1^n(I_{nk}))+l(P_2^n(I_{nk})) \right]. 
\end{equation}
The feature extraction network parameters are learnt by minimizing $\mathcal{L}$. We employ ResNet-50 as the feature extraction network in our experiments. We learn contrastive features for each subband of a Laplacian pyramid decomposition by feeding the subbands instead of the images to the above contrastive learning framework. We learn network parameters independently for each subband. 

\subsection{Blind Quality Prediction}
    We design a quality prediction model based on our features obtained through contrastive learning using the framework of NIQE \cite{DBLP:journals/spl/MittalSB13}.
    We extract features from $M$ subbands of an $M$-level Laplacian pyramid decomposition. Let the features obtained for Subband-$m$, $m\in\{1,2,\cdots,M\}$ be $\mathbf{z}_m$. Let $\mathbf{z}_0$ denote the features obtained from the  image directly. We concatenate all these features to obtain $(\mathbf{z}_0,\mathbf{z}_1,\cdots,\mathbf{z}_M)$. However, the dimension of this resulting feature vector is very large. Thus, we perform principal component analysis (PCA) to reduce the feature dimension to $D$ components for fair comparison with other features. We then use these reduced dimensionality features from the test image to compute quality as follows. 
    
    Similar to the NIQE evaluation framework, we take a corpus of pristine image patches of size $P\times P$ satisfying the sharpness criterion described in  \cite{DBLP:journals/spl/MittalSB13}. 
    We select these patches from scenes that do not overlap with any of the test images.  
    %We use $Q= 96$ and the sharpness threshold equal to 0.3 in our experiments.
    Since low light restored images also suffer from color artifacts, we believe that choosing colorful pristine patches can help better assess the quality of the restored images. Thus, among the selected sharp patches, we further compute the colorfulness of each patch using the method described in \cite{10.1117/12.477378}. We select those patches possessing a colorfulness index greater than a threshold $\tau_c$. We obtain the features from the sharp and colorful pristine patches by applying our feature extraction network and subject them to PCA. 
     Let the pristine multivariate Gaussian (MVG) model  learnt on the reduced dimensionality features be $(\mu_{r},\Sigma_{r})$.
     For each test image, we take all $P\times P$ patches with an overlap of $P/2$, extract features, reduce dimensionality and fit an MVG model over them to determine $(\mu_{d},\Sigma_{d})$.
    Finally we compute the quality score of the test image as \begin{equation}
    Q=\sqrt{\left(\left(\mu_{r}-\mu_{d}\right)^{T}\left(\frac{\Sigma_{r}+\Sigma_{d}}{2}\right)^{-1}\left(\mu_{r}-\mu_{d}\right)\right)}.
    \label{eq:mvg_distance}
    \end{equation}
    
\section{Experiments}

\subsection{Performance Evaluation and Benchmarking}
We benchmark state of the art full reference (FR) and supervised and unsupervised NR IQA measures on our dataset. Popular FR measures that we benchmark include PSNR, SSIM \cite{DBLP:journals/tip/WangBSS04}, Feature SIM \cite{DBLP:journals/tip/ZhangZMZ11} and Multiscale SSIM \cite{1292216}. We also benchmark RIQMC \cite{7056527} which is a contrast distortion based FR measure. Finally we benchmark LPIPS \cite{DBLP:journals/corr/abs-1801-03924}, which is a deep learning based full reference measure. We use the high exposure shot of the scene as the reference for computing the full reference quality scores.

We benchmark no reference measures that need to be trained against human opinion scores and refer to them as supervised NR IQA measures. Among non-deep learning based methods, we evaluate BRISQUE \cite{6272356}, PI \cite{PI} and Ma \etal \cite{DBLP:journals/corr/MaYY016}.
We also compare recent deep learning based NR IQA methods such as MetaIQA \cite{9156932} and NIMA \cite{Talebi2018NIMANI}. Finally, we evaluate a simple benchmark based on pretrained ResNet-50 features obtained after global average pooling using a support vector regression model as in \cite{8103112}.  

Among unsupervised IQA measures we compare against  NIQE \cite{DBLP:journals/spl/MittalSB13} and IL-NIQE \cite{7094273}. Both these methods require a set of pristine images for comparison and we use the same set as that which is used for our prediction framework as described in the implementation details.
We also benchmark NIQMC \cite{DBLP:journals/corr/abs-1904-08879} which is an unsupervised contrast distortion based blind quality index.

\textbf{Implementation details of M-SCQALE:} We learn quality features through contrastive learning using images from a larger corpus of 60,000 distorted images that we create similar to DSLR. In particular, we take 300 scenes from the training set of the SID dataset, different from the test set of SID that is used in creating DSLR. We then apply contrast enhancement and denoising algorithms on these images similar to DSLR. Thus, there is no overlap of scene content between the images on which contrastive learning is performed and DSLR.  After feature learning, we use high exposure/pristine images from the SID \cite{SID} training set (300 images) and LOL dataset \cite{DBLP:conf/bmvc/WeiWY018} (500 images) to learn the feature statistics for our pristine MVG model.

  We choose the number of scales as $M = 3$ in our experiments and study the performance variation with $M$ in Section \ref{sec:ablations}. We use two RTX 2080 Ti GPUs using Pytorch framework for our training.
    Starting at the image level, we choose the number of scenes as $N = 4$ and the number of distorted versions $K = 10$ in a mini-batch. As the resolution of the subbands decreases with scale, based on our GPU memory constraints, we increase $N$ and $K$ by a factor of 2 in each scale. The exact values for each sub-band can be found in the supplementary. 
    We train a ResNet-50 network for each sub-band independently for 110 epochs using Adam optimizer with a learning rate of 0.01.
    We set the temperature parameter as $\tau =  0.1$ in all our experiments.
    We use sharpness threshold as $\tau_{s} = 0.3$ times the maximum sharpness and colorfulness threshold $\tau_{c} = 0.8$ times the maximum colorfulness in each image for selecting the pristine patches in our quality prediction framework. We set $D = 2048$ and $P = 96$ similar to NIQE \cite{DBLP:journals/spl/MittalSB13}.

\textbf{Performance evaluation: }
 We use Spearman's rank order correlation coefficient (SRCC) and Pearson's linear correlation coefficient (PLCC) between the subjective scores and quality predictions to evaluate performance. PLCC is computed after passing the predicted scores through a non-linearity as described in \cite{1709988}.
 For all methods which require training, we randomly split the dataset into 80\% for training and 20\% for testing such that there is no scene overlap between the train and the test splits. We report the median performance across 100 such splits in Table \ref{tab:medianperformance}. For fair comparison among all the methods, we evaluate the methods which do not require training also on the 20\% test splits and report their median performance. 

 \begin{table}
  \centering
  \begin{tabular}{ |c|c|c|c|c|}
% Method & SRCC   & PLCC \\
\hline
Method & \multicolumn{2}{c|}{SRCC}  & \multicolumn{2}{c|}{PLCC}\\
\hline
& Median & Std & Median & Std \\
\hline
\multicolumn{5}{|c|}{Full Reference measures} \\
\hline
PSNR & 0.39  & 0.05 & 0.41&0.04\\
RIQMC \cite{7056527} & 0.16 & 0.07 &0.22 &0.06\\
SSIM \cite{DBLP:journals/tip/WangBSS04} &  0.66 & 0.04 & 0.68 & 0.03  \\
FSIM \cite{DBLP:journals/tip/ZhangZMZ11} &  0.72 & 0.03 &0.74 & 0.03 \\
MS-SSIM \cite{1292216}    &0.72& 0.03 & 0.74&0.03 \\
LPIPS \cite{DBLP:journals/corr/abs-1801-03924} &\textbf{0.83}&0.03 &\textbf{0.82}&0.02 \\
\hline
\multicolumn{5}{|c|}{Supervised No Reference measures} \\
\hline
BRISQUE \cite{6272356} & 0.57 & 0.04 & 0.61&0.03 \\
% Hypernet & & && \\
Meta-IQA \cite{Zhu2020MetaIQA} & 0.80 &0.01&0.73&0.02\\
NIMA \cite{Talebi2018NIMANI }&0.78&0.03&0.78&0.03\\
Ma \etal \cite{DBLP:journals/corr/MaYY016} & 0.53 &0.04&0.60&0.04\\
PI \cite{PI} & 0.53 &0.04 &0.59 &0.04\\
Resnet50 + SVR & \textbf{0.84} & 0.03&\textbf{0.84}&0.03\\
\hline
\multicolumn{5}{|c|}{Unsupervised No Reference measures} \\
\hline
NIQMC \cite{DBLP:journals/corr/abs-1904-08879} & 0.17 &0.06&0.22&0.06 \\
NIQE \cite{DBLP:journals/spl/MittalSB13} &0.54& 0.04&0.56&0.04\\
IL-NIQE \cite{7094273} &0.50&0.05&0.49&0.08 \\
M-SCQALE    &  \textbf{0.70} & 0.03 &\textbf{0.71} &0.03 \\
\hline
\end{tabular}
  \caption{Median performance on 100 random train test splits of DSLR dataset. Std refers to the standard deviation in performance.}
  \label{tab:medianperformance}
\end{table}

\textbf{Performance analysis: } We observe that among the FR measures, LPIPS achieves the best performance. 
%Although LPIPS is being increasingly used to optimize low light restoration methods, our benchmarking reinforces its superior performance. 
Among supervised NR measures, we see that ResNet-50 features that have been pre-trained for image classification can be regressed quite well to predict MOS. This is also not surprising given its stable competitive performance on multiple recent authentically distorted IQA datasets \cite{DBLP:journals/corr/abs-1912-10088,Ghadiyaram2016MassiveOC}. Among the unsupervised NR IQA models, M-SCQALE achieves the state of the art performance. This shows that contrastive learning can be effectively used to learn features that can predict quality in an unsupervised fashion.

\subsection{Ablations}\label{sec:ablations}
We evaluate the strength of different components of M-SCQALE. Various experiments in this subsection involve comparing our features and feature learning method with other feature extraction mechanisms. We perform all these comparisons by replacing our features with other features in Equation (\ref{eq:mvg_distance}) for unsupervised NR QA. As M-SCQALE is training free, we report the SRCC between the predicted scores and the actual MOS on the entire DSLR dataset. 

\textbf{Other deep features vs. M-SCQALE:} We experiment with  various self-supervised pretrained ResNet-50 models such as
Simclrv2 \cite{DBLP:journals/corr/abs-2006-10029}, 
Mocov2 \cite{DBLP:journals/corr/abs-2003-04297}, 
CMC \cite{DBLP:journals/corr/abs-1906-05849} and
Swav \cite{DBLP:journals/corr/abs-2006-09882} to extract features and use them in our quality prediction framework. We also compare with features extracted  with a ResNet-50 that has been pre-trained in a supervised fashion for image classification denoted as Sup* ResNet-50. We see from Table \ref{tab:featurecomparisons} that our features outperform these pre-trained features although they have been trained on huge amounts of data and with far higher computing power. Thus our contrastive learning paradigm contributes to this performance difference with other deep features. Further, we also try training the frameworks of Simclr \cite{DBLP:journals/corr/abs-2002-05709} and CMC \cite{DBLP:journals/corr/abs-1906-05849} on our dataset. However, they perform poorer than the pre-trained features, perhaps due to the lack of enough training data. Exact results can be found in the supplementary.
\begin{table}[]
\centering
\begin{tabular}{@{}cc@{}}
\toprule
Features            & SRCC                                 \\ \midrule
Sup* ResNet-50 & 0.46                                 \\
Simclrv2 \cite{DBLP:journals/corr/abs-2006-10029}     & 0.40                                 \\
Mocov2 \cite{DBLP:journals/corr/abs-2003-04297}       & 0.50\\
CMC \cite{DBLP:journals/corr/abs-1906-05849}          & 0.41                                 \\
Swav \cite{DBLP:journals/corr/abs-2006-09882}         & 0.47                                 \\
M-SCQALE&  \textbf{0.70} \\ \bottomrule
\end{tabular}
\caption{Evaluating various self-supervised features.}
  \label{tab:featurecomparisons}
\end{table}
\textbf{Choice of patches from the same scene for M-SCQALE:}
One of the key aspects of our training method is the way we choose patches for multiple views. In particular, we choose pairs of patches such that the negative pairs come from different distorted versions of the same scene. We believe that this is critical in learning features that relate to quality and not content. To understand the benefit of this aspect, we experiment with a learning method where the negative pairs can come from images of different scenes. We perform this comparison for the features learnt for a single scale (just the image). The first two rows in Table \ref{tab:strengthofdifferentcomponents} reveal a huge difference in the performance indicating the importance of scene content remaining the same while computing the contrastive loss. 

\begin{figure}
        \centering
              \includegraphics[width=\linewidth]{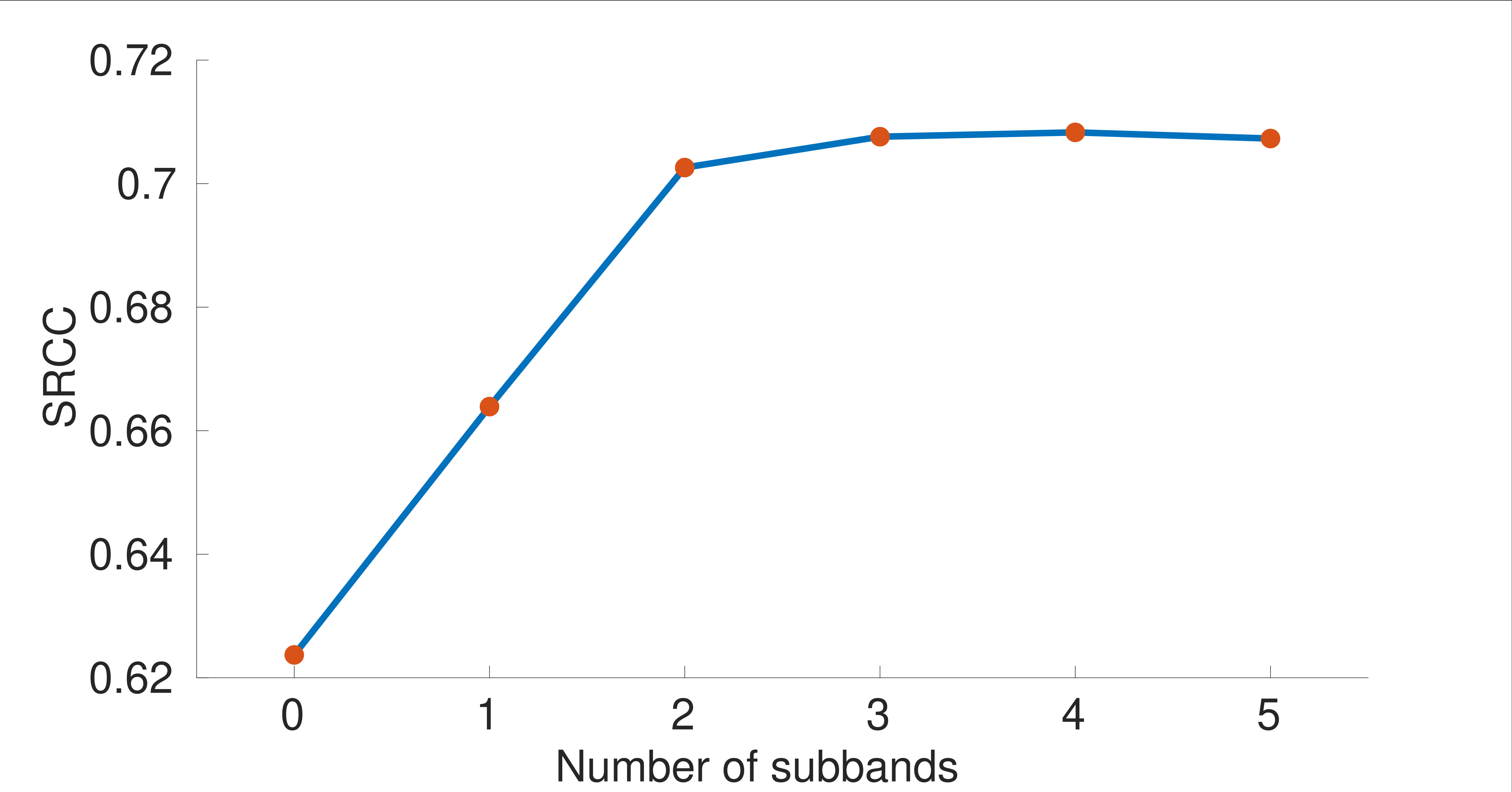}
              \caption{Performance variation of M-SCQALE for different number of sub-bands.}
              \label{fig:no_of_subbands}
        \end{figure}

\textbf{Multiscale subband features:} Recall that in M-SCQALE we evaluate the features in different sub-bands of a Laplacian pyramid decomposition and concatenate all these features learnt along with the features extracted from the image. We study the performance variation with the number of scales in the Laplacian pyramid decomposition $M$ and report the performance in Figure \ref{fig:no_of_subbands}. $M = 0$ implies features are extracted only at the image level. $M = 1$ implies feature extraction at the image level, the first high pass sub-band and the corresponding low pass band and so on for other values of $M$. We observe that after $M = 3$, the performance saturates. We also show the comparison of the multi-scale feature extraction method for $M=3$ in Table \ref{tab:strengthofdifferentcomponents}. 
       
\textbf{Role of sharpness and colorfulness:}
We next analyze the strength of the different criteria we use for patch selection from the pristine patches to learn the pristine MVG model. We evaluate the impact of patch selection using both sharpness and colorfulness. We see from Table \ref{tab:strengthofdifferentcomponents} that selecting colorful patches improves the performance. 

\begin{table}
    \centering
    \begin{tabular}{|c|c|c|c|c|}
    \hline
         Same   &Multi-&Sharp-&Colorful-&SRCC\\
         scene&scale&ness&ness&\\
         patches&&&&\\
         
    \hline
    \xmark& \xmark&\cmark & \cmark& 0.13 \\
    \hline
    \cmark& \xmark&\cmark & \cmark& 0.62 \\
    \hline
    \cmark& \cmark&\xmark & \xmark& 0.64 \\
    \cmark& \cmark&\cmark & \xmark& 0.66 \\
    \hline
    \cmark& \cmark&\cmark & \cmark& 0.70 \\
    \hline
    
    \end{tabular}
    \caption{Strength of different components of our method.}
    \label{tab:strengthofdifferentcomponents}
\end{table}

\subsection{Performance analysis on other datasets}
We evaluate M-SCQALE on other datasets containing low light images such as CID \cite{DBLP:journals/tip/VirtanenNVOH15} and NNID \cite{DBLP:journals/tmm/XiangYG20}. We do not learn features using images from these datasets and use the same model evaluated on DSLR. Thus, there is no training of any form on these datasets. As in \cite{DBLP:journals/tmm/XiangYG20}, we evaluate M-SCQALE on the 79 low-contrast images in CID. NNID contains 2240 images with distortions such as under-enhancement, blur and color distortion. We evaluate different unsupervised models on the above datasets in Table \ref{tab:otherdatasets}. 
We observe that NNID contains images with widely varying resolutions. To be consistent with the resizing step of IL-NIQE, we resize all the images to a resolution of $512\times 512$ before evaluating NIQE and M-SCQALE on NNID. This resizing is important especially since these approaches use fixed patch sizes. We see that M-SCQALE achieves the best performance on both these datasets demonstrating its excellent generalization performance. 

\begin{table}
\centering
\begin{tabular}{ccccc}
\toprule
Dataset & \multicolumn{2}{c}{CID \cite{DBLP:journals/tip/VirtanenNVOH15}} & \multicolumn{2}{c}{NNID\cite{DBLP:journals/tmm/XiangYG20}} \\
\cmidrule(r){2-3}\cmidrule(l){4-5}
Method & {SRCC} & {PLCC}  & {SRCC} & {PLCC} \\
\midrule
% NIQE\cite{DBLP:journals/spl/MittalSB13} & 0.68 & 0.69 & 0.58 & 0.57\\

NIQE\cite{DBLP:journals/spl/MittalSB13} & 0.68 & 0.69 & 0.61 & 0.61\\
IL-NIQE\cite{7094273} & 0.54 & 0.62 & 0.68 & 0.68\\
M-SCQALE &\textbf{0.87}& \textbf{0.89} & \textbf{0.77} &\textbf{0.77} \\

\bottomrule
\end{tabular}
\caption{Performance analysis on other datasets.}
\label{tab:otherdatasets}
\end{table}

\subsection{Visualization of features}
\begin{figure}
        \centering
          \includegraphics[width=\linewidth]{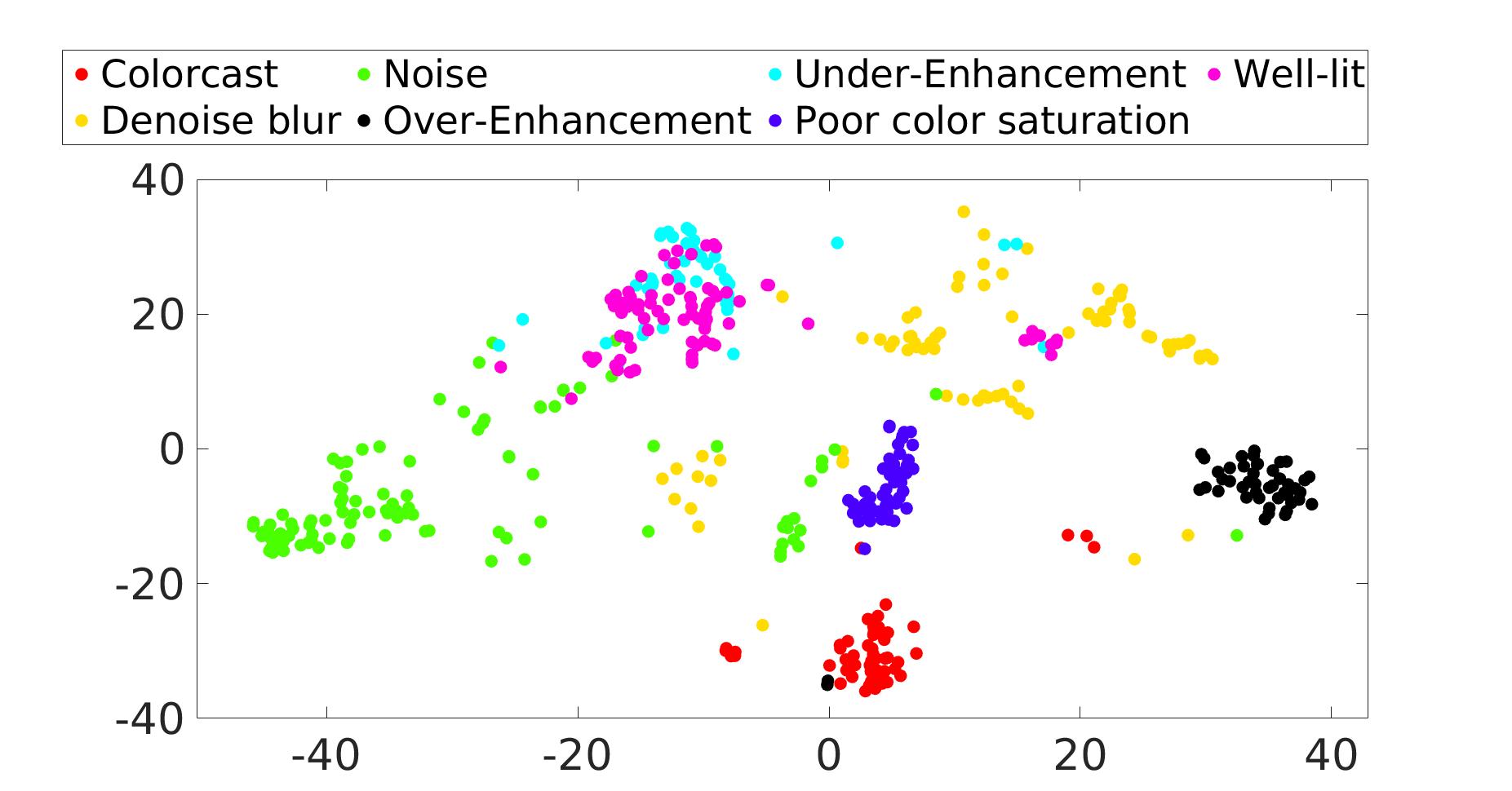}
              \caption{Scatter plot of 2-dimensional features obtained by using t-SNE on M-SCQALE features.}
              \label{fig:tsneplot}
        \end{figure}

We also try to visually interpret the M-SCQALE features using a t-SNE plot. We compute features on all images of DSLR, then reduce dimensionality to 50 dimensions using PCA following \cite{Maaten2008VisualizingDU}, and then visualize the features. We show the scatter plot of t-SNE features from different distortion types in  Figure \ref{fig:tsneplot}. We observe that the different distortion types form clearly demarcated clusters. This explains that our contrastively learnt features are able to differentiate images in the distortion space.   

\begin{figure}
\centering
\begin{minipage}{0.45\linewidth}
\centering
\subcaptionbox{(a) Predicted = 57.17, MOS = 26.71}
{\includegraphics[width=\textwidth]{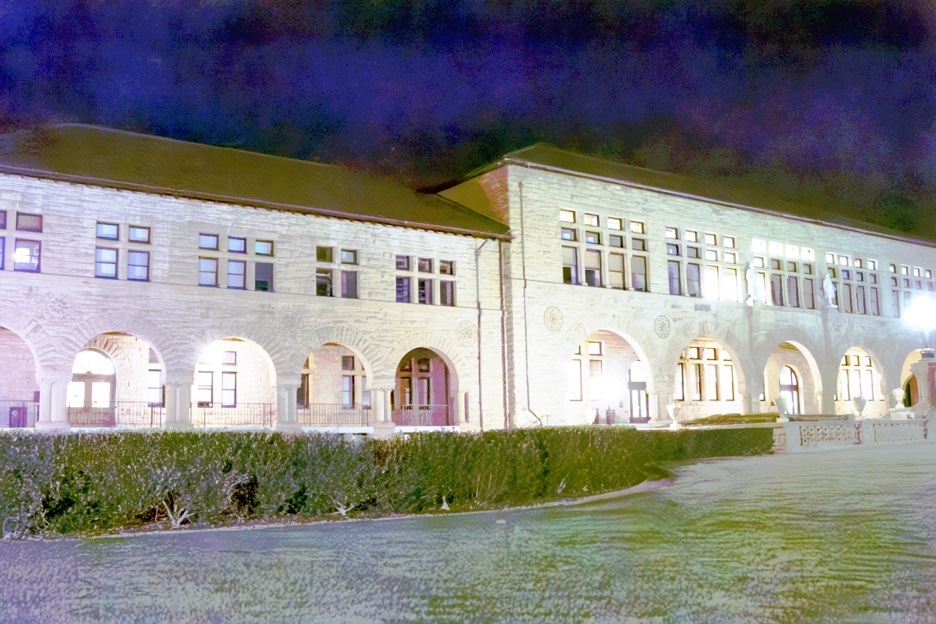}}

\end{minipage}
\hspace{0.12cm}
\begin{minipage}{0.45\linewidth}
\centering
\subcaptionbox{(b) Predicted = 41.60, MOS = 61.86}
{
\includegraphics[width=\textwidth]{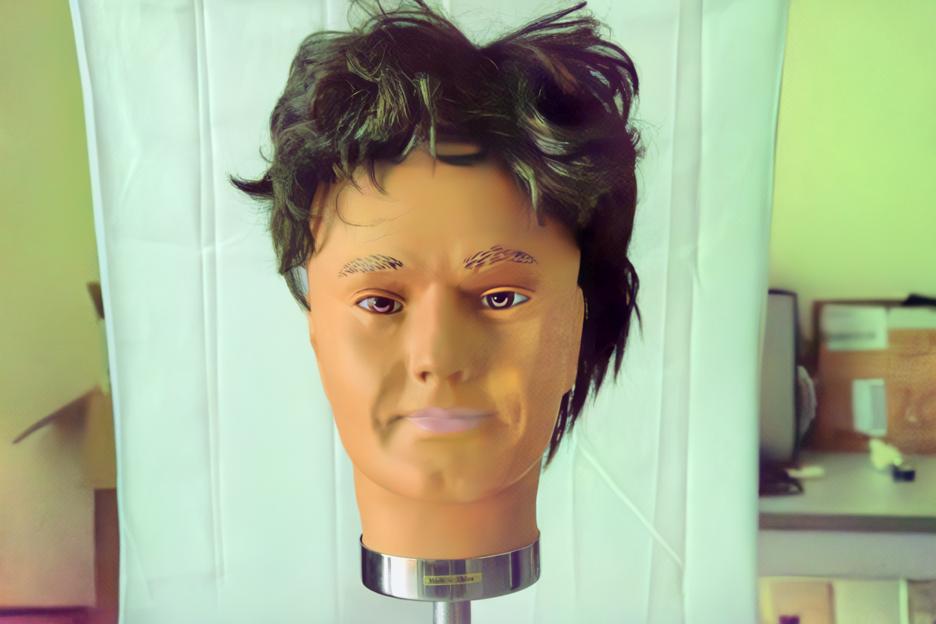}}

% \caption{(b)}
\end{minipage}

\caption{Failure cases.}
\label{fig:failurecases}
\end{figure}

\textbf{Failure cases:} We analyze where M-SCQALE fails by comparing the subjective scores and our quality predictions passed through a non-linearity to predict MOS. As shown in Figure \ref{fig:failurecases} (a), M-SCQALE overestimates the quality of over enhanced images. Although our feature learning mechanism is able to identify such distortions, we believe that our unsupervised quality prediction framework still lacks the artillery to carefully quantify such distortions. 
In another example shown in Figure \ref{fig:failurecases} (b), we observe that our model underestimates the quality of this colorful image. This could be happening because the distribution of colorful pristine patches may not be as colorful as the enhanced image. 

\section{Conclusion}
Our DSLR database represents one of the first exhaustive QA studies on low light restored images. Our novel unsupervised NR method for QA of low light restored images, M-SCQALE, addresses the twin challenges of the need for a reference image as well as human quality labels for training. In particular, we show that contrastive learning can be a valuable tool for effective learning of quality aware features. Although, we do not train on subjective data, M-SCQALE still achieves very good performance in accordance with human judgements. While the features we learn are able to distinguish distortion types, their unsupervised mapping to perceptual quality can be improved further.

\setcounter{section}{0}
\part*{Supplementary Material}
\noindent This supplement includes the following:

\begin{enumerate}
       
\item Low Light Restoration (LLR) Algorithms used to generate DSLR.
\item Block Diagram of M-SCQALE.
\item Additional Training Details.
\item Additional Experiment Details.
\item Additional Details of Subjective Study.
    \end{enumerate}

\section{LLR Algorithms used to generate DSLR}
 In Table \ref{tab:resttechs}, we report the list of LLR algorithms we use to generate images in DSLR. We use a combination of 14 contrast enhancement techniques coupled with 3 denoisers, and also 7 joint contrast enhancement and denoising techniques which include deep learning methods. Note that we use a variety of algorithms belonging to different categories such as histogram equalization based, retinex based and multi-scale approaches. 
 \begin{table*}
 \centering
 \begin{tabular}{|c|}
      \hline
     \textbf{Contrast Enhancement Techniques}\\
     \hline 
     Automatic Color Enhancement (ACE) and its Fast Implementation \cite{ACE}\\
     \hline
     Efficient Contrast Enhancement Using Adaptive Gamma Correction With Weighting Distribution \cite{AGCWD}\\
     \hline
     A Bio-Inspired Multi-Exposure Fusion Framework for Low-light Image Enhancement \cite{BIMEF}\\
     \hline
A New Low-Light Image Enhancement Algorithm Using Camera Response Model \cite{CAMERAMODEL}\\
\hline
Contrast-limited adaptive histogram equalization \cite{CLAHE}\\
\hline
Contextual and Variational Contrast Enhancement \cite{CVC}\\
\hline
Single Image Haze Removal Using Dark Channel Prior \cite{DCP}\\
\hline
A New Image Contrast Enhancement Algorithm Using Exposure Fusion Framework \cite{EFF}\\
\hline
A fusion-based enhancing method for weakly illuminated images \cite{FUSE}\\
\hline
Histogram Equalization\\
\hline
Contrast Enhancement Based on Layered Difference Representation of 2D Histograms \cite{LDR}\\
\hline
LIME: Low-Light Image Enhancement via Illumination Map Estimation \cite{LIME}\\
\hline
Naturalness Preserved Image Enhancement Using a Priori Multi-Layer Lightness Statistics \cite{NPEAMLLS}\\
\hline
Weighted Adaptive Histogram Equalization \cite{WAHE}\\
\hline
\textbf{Joint Contrast Enhancement and Denoising Techniques}\\
     \hline
     Fast and Efficient Image Quality Enhancement via Desubpixel Convolutional Neural Networks
     \cite{FEQE}\\
     \hline
     MBLLEN: Low-Light Image/Video Enhancement Using CNNs \cite{MBLLEN}\\
     
     \hline
LLRNET \cite{LLRNET} \\
\hline
UNET \cite{UNET} \\
\hline
Learning to See in the Dark \cite{SID}\\
\hline
Joint Enhancement and Denoising Method via Sequential Decomposition \cite{JED}\\
\hline
Structure-Revealing Low-Light Image Enhancement Via Robust Retinex Model \cite{Robustretinex}\\
\hline
\textbf{Denoisers}\\
     \hline
     Deep Iterative Down-Up CNN for Image Denoising (DIDN)
     \cite{DIDN}\\
     \hline
    DIDN pre-trained \cite{DIDN}\\
     
     \hline
     The Noise Clinic: a Blind Image Denoising Algorithm \cite{noiseclinic}\\
\hline
 \end{tabular}
 \caption{List of low light restoration algorithms used to generate DSLR.}
 \label{tab:resttechs}
\end{table*}

\section{Block Diagram of M-SCQALE}
\begin{figure}[h]
     \centering
     \includegraphics[width=\linewidth]{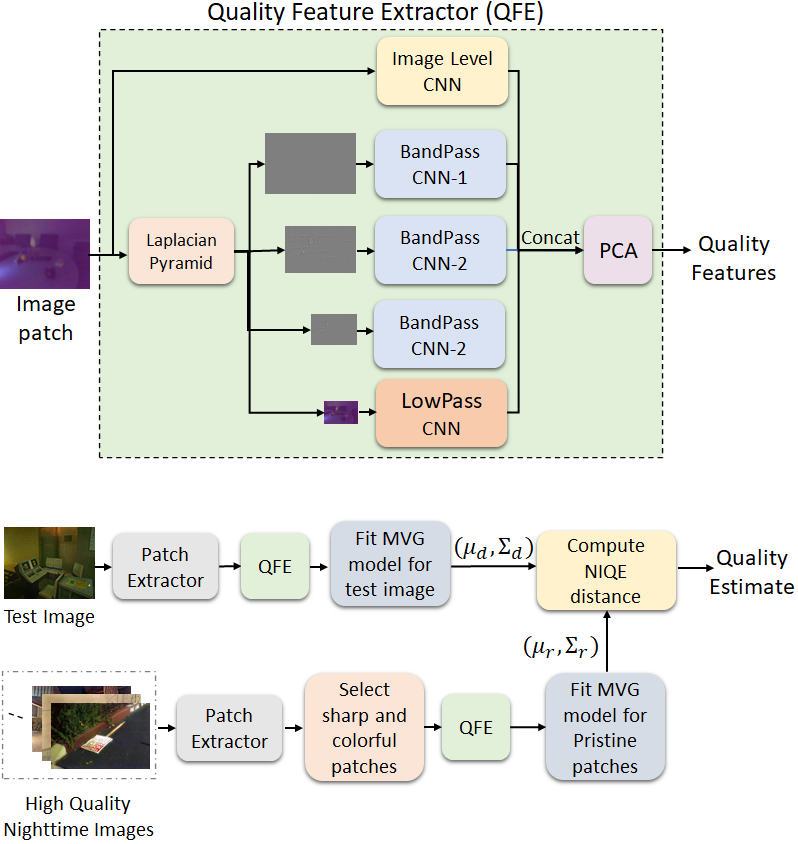}
     \caption{Block diagram of M-SCQALE.}
     \label{fig:blockdiagram}
 \end{figure}

In Figure \ref{fig:blockdiagram}, we show the complete workflow of the our M-SCQALE method for unsupervised NR-QA of restored low light images. Note that NIQE distance refers to Equation (3) in the main paper.

\section{Additional Training Details}
\subsection{Batch sizes}For training different sub-bands in M-SCQALE, we choose the number of scenes $N$ and the number of distorted versions $K$ in a mini-batch as given in Table \ref{tab:noofsubbands}.
\begin{table}
     \centering
     \begin{tabular}{|c|c|c|}
     \hline
          \textbf{Level} & $\mathbf{N}$ & $\mathbf{K}$ \\
     \hline
          Image level & 4 & 10\\
     \hline
          Level 1 high pass sub-band & 4 & 10\\
     \hline
          Level 2 high pass sub-band & 8 & 20\\
     \hline
          Level 3 high pass sub-band  & 16 & 40\\
     \hline
         Low pass sub-band & 32 & 40\\
     \hline
          
     \end{tabular}
     \caption{Values of $N$ and $K$ used for different sub-bands.}
     \label{tab:noofsubbands}
 \end{table}
\section{Additional Experiment Details}
\subsection{Training Simclr and CMC for Quality Assessment}
In the main paper, we reported the performance of using features from pre-trained Simclr and CMC ResNet-50 networks in our quality prediction framework. We remarked there that training these frameworks on the set of low light restored images on which M-SCQALE's contrastive learning framework was trained, results in poorer performance as compared to the pre-trained features. We report the performance in Table \ref{tab:simclr&cmc}.

For Simclr, we follow the exact method proposed by Chen \etal \cite{simclr}. In a minibatch, $B$ data points are randomly sampled and their corresponding augmented pairs are generated using the prescribed augmentation procedure resulting in $2B$ data points. Given a sample and it's augmented version (positive pair), the remaining $2(B-1)$ data points in the minibatch are treated as negative examples.

Similarly for CMC, we follow the exact method proposed by Tian \etal \cite{CMC}. During training, Y and DbDr from the
same data point are considered as the positive pair, while DbDr channels
from other randomly sampled data points are considered as a negative pair (for a given Y).

We use two RTX 2080 Ti GPUs using Pytorch framework for our training. We train a ResNet-50 network for each framework for 110 epochs using Adam optimizer with a learning rate of 0.01.
We set the temperature parameter as $\tau =  0.1$. We observe that setting $\tau = 0.07$ for CMC (as in the original CMC paper) gives very poor performance for our task and hence we report the performance with $\tau = 0.1$. We use a batch size of 40 as limited by our memory constraints. We use a patch size of 312 as used in M-SCQALE's image level contrastive learning framework for fair comparison.

\begin{table}
     \centering
     \begin{tabular}{|c|c|}
     \hline
          \textbf{Method} & \textbf{SRCC}  \\
     \hline
          Simclr & 0.13 \\
    
     \hline
          CMC & 0.28\\ 
     \hline
     \end{tabular}
     \caption{Results of training Simclr and CMC for quality assessment.}
     \label{tab:simclr&cmc}
 \end{table}
 
 \subsection{Unsupervised IQA with CORNIA:}
 CORNIA \cite{CORNIA} is an example of an unsupervised feature learning algorithm. Typically, CORNIA features are used in a supervised setting, where the features are trained against human opinion scores. However, to compare these features against M-SCQALE's features, we design an unsupervised quality index based on the CORNIA features. We use CORNIA features instead of M-SCQALE's features in our NIQE \cite{NIQE} based quality prediction framework.
 
  In our experiment, we first learn the dictionary in an unsupervised fashion \cite{CORNIA} on the set of low light restored images on which M-SCQALE's contrastive learning framework was trained. We then use the learnt features from in our quality prediction framework. This gives a median SROCC of 0.25 and a median PLCC of 0.23 on the same 100 test splits as reported in Table 1 in the main paper.
  
 \section{Additional Details of Subjective Study}
\begin{figure}
\centering
\begin{minipage}[b]{1\linewidth}
\centering
\subcaptionbox{Influence of resolution.}
{\includegraphics[width=\textwidth]{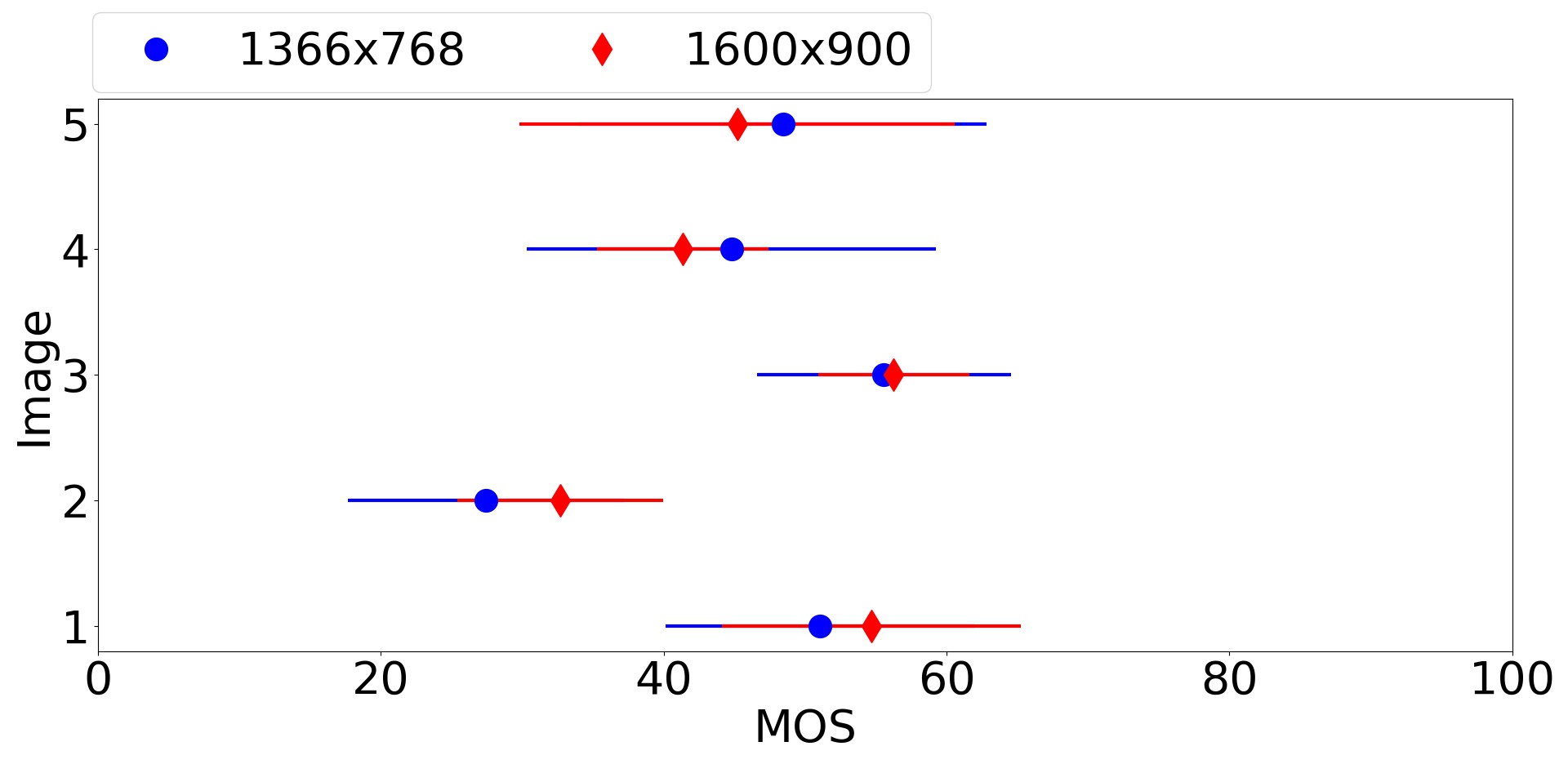}}
\end{minipage}
\\
\begin{minipage}[b]{1\linewidth}
\centering
\subcaptionbox{Influence of subject viewing distance.}
{\includegraphics[width=\textwidth]{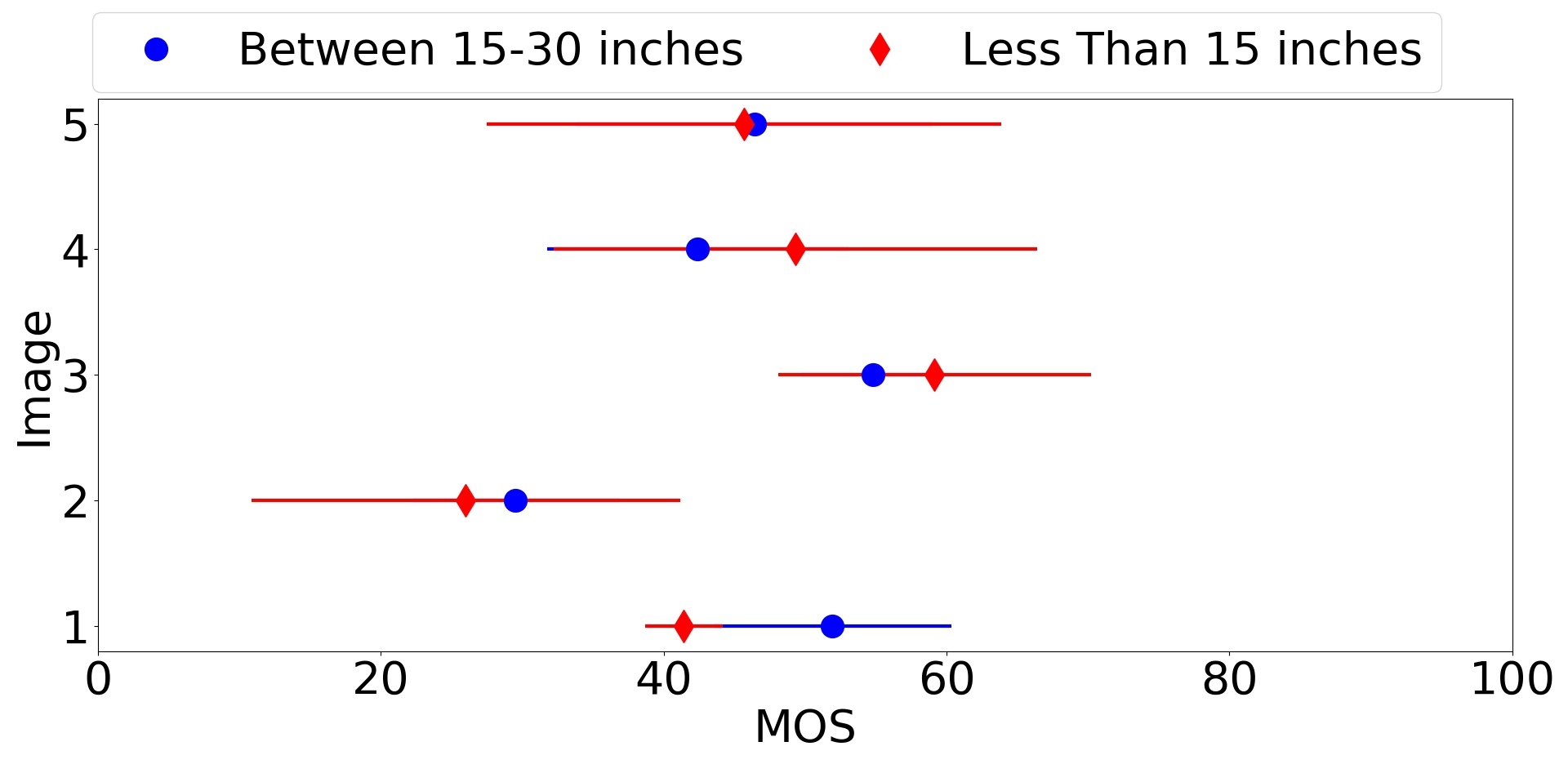}}
\end{minipage}
\\
\begin{minipage}[b]{1\linewidth}
\centering
\subcaptionbox{Influence of screen size.}{
\includegraphics[width=\textwidth]{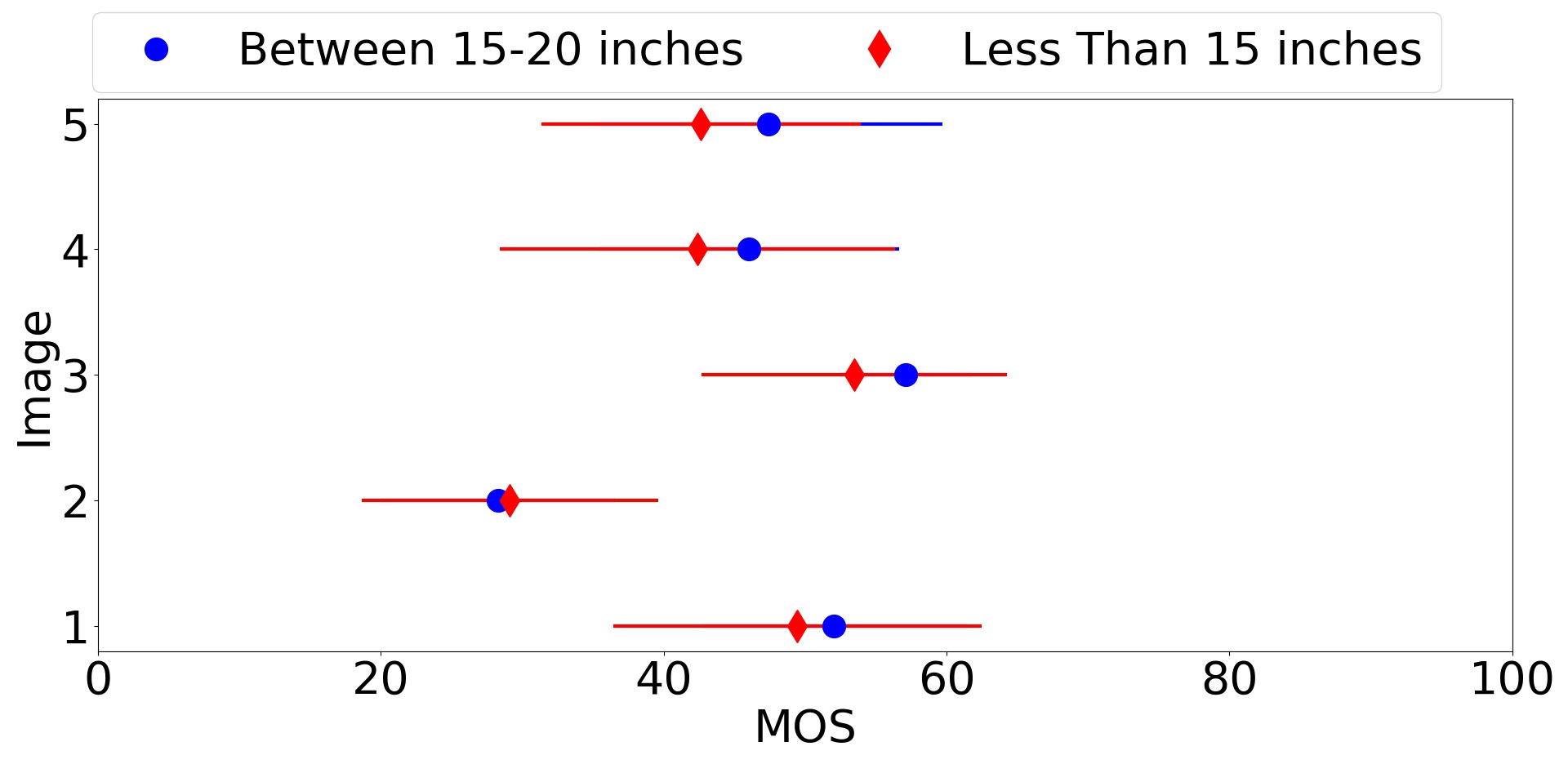}}
\end{minipage}
\caption{Plots illustrating the influence of various parameters on a subject's quality rating on five randomly sampled images from DSLR. }
\label{fig:studyparams}
\end{figure}

 We recorded various parameters such as screen size, screen resolution, subject distance from screen using a mandatory survey form at the start of the subjective study. We study the role of these parameters in a similar fashion as analyzed by Deepti \etal in \cite{Ghadiyaram2016MassiveOC}. We compute the MOS from disjoint sections of the population separated based on these parameters for 5 randomly sampled images from DSLR. We plot the corresponding MOS values and their associated 95\% confidence intervals in Figure \ref{fig:studyparams}. We observe that the difference between the mean of the ratings obtained on the five randomly sampled image between disjoint sets of the population separated by either resolution, viewing distance or screen size is not statistically significant. However we do not disregard the plausible influences that these parameters might have on distortion perception based on an analysis of five randomly sampled images. A study that exactly studies the relationship between a subject's perception of quality and the inter-play of these factors is an interesting area for future work, considering that crowdsourced studies are becoming very popular in recent times. 
%%%%%%%%% REFERENCES
{\small
\bibliographystyle{ieee_fullname}
\bibliography{refs}
}
\end{document}